\documentstyle[12pt,twoside]{article}

\input FEYNMAN

\def\ie{\hbox{i.e.}{}} 		
\def\eg{\hbox{e.g.}{}} 		\def\cf{\hbox{cf.}{}}
\def\etal{\hbox{et al.}{}} 	
\def\ps{\hbox{p.s.}{}}
\def\cm{\hbox{c.m.}{}}

\def\ltap{\;\raisebox{-.5ex}{\rlap{$\sim$}} \raisebox{.5ex}{$<$}\;}
\def\gtap{\;\raisebox{-.5ex}{\rlap{$\sim$}} \raisebox{.5ex}{$>$}\;}
\def\r#1{\ignorespaces $^{#1}$}

\def\beq{\begin{equation}}
\def\eeq{\end{equation}}

\def\beqn{\begin{eqnarray}}
\def\eeqn{\end{eqnarray}}

\makeatletter

\def\section{\@startsection{section}{1}{\z@}{3.5ex plus 1ex minus
   .2ex}{2.3ex plus .2ex}{\large\bf}}

\def\thesection{\arabic{section}.}

\def\appendix{\setcounter{section}{0}
 \def\thesection{APPendIX \Alph{section}:}
 \def\theequation{\Alph{section}.\arabic{equation}}}

\@addtoreset{equation}{section}
\def\theequation{\thesection\arabic{equation}}

\def\@normalsize{\@setsize\normalsize{15pt}\xiipt\@xiipt
\abovedisplayskip 14pt plus3pt minus3pt%
\belowdisplayskip \abovedisplayskip
\abovedisplayshortskip \z@ plus3pt%
\belowdisplayshortskip 7pt plus3.5pt minus0pt}

\def\small{\@setsize\small{13.6pt}\xipt\@xipt
\abovedisplayskip 13pt plus3pt minus3pt%
\belowdisplayskip \abovedisplayskip
\abovedisplayshortskip \z@ plus3pt%
\belowdisplayshortskip 7pt plus3.5pt minus0pt
\def\@listi{\parsep 4.5pt plus 2pt minus 1pt
     \itemsep \parsep
     \topsep 9pt plus 3pt minus 3pt}}

\makeatother

\def\pl#1#2#3{Phys. Lett. {\bf #1} (19#2) #3}

\def\prl#1#2#3{Phys. Rev. Lett. {\bf #1} (19#2) #3}

\def\prev#1#2#3{Phys. Rev. {\bf #1} (19#2) #3}
\def\np#1#2#3{Nucl. Phys. {\bf #1} (19#2) #3}
\def\app#1#2#3{Acta Phys. Pol. {\bf #1} (19#2) #3}
\def\apa#1#2#3{Acta Phys. Austr. {\bf #1} (19#2) #3}
\def\jcp#1#2#3{J. Comp. Phys. {\bf #1} (19#2) #3}
\def\ptp#1#2#3{Prog. Theor. Phys. {\bf #1} (19#2) #3}

\begin{document}

%
\newcommand{\sss}{\scriptscriptstyle}
\newcommand{\rar}{\rightarrow}
\renewcommand{\to}{\rar}

\newcommand{\elp}{$e^{\sss +}$}
\newcommand{\elm}{$e^{\sss -}$}
\newcommand{\elpm}{$e^{\sss \pm}$}
\newcommand{\epem}{$e^{\sss +}e^{\sss -}$}
\newcommand{\Z}{$Z^{\sss 0}$}
\newcommand{\Wpm}{$W^{\sss \pm}$}
\newcommand{\Wp}{$W^{\sss +}$}
\newcommand{\Wm}{$W^{\sss -}$}
\newcommand{\ve}{$\nu_e$}
\renewcommand{\tt}{$\bar{t}t$}
\newcommand{\ttb}{\bar{t}t}

\newcommand{\mt}{m_t}
\newcommand{\mz}{M_{\sss Z}}
\newcommand{\mw}{M_{\sss W}}
\newcommand{\mb}{m_{\sss b}}
\newcommand{\mwb}{M_{\sss Wb}}

\newcommand{\rs}{\sqrt{s}}
\newcommand{\costop}{\cos \theta_{\sss t}}
\newcommand{\wt}{\Gamma_t}
\newcommand{\ww}{\Gamma_{\sss W}}
\newcommand{\wz}{\Gamma_{\sss Z}}
\newcommand{\sth}{\sin\theta_{\sss W}}
\newcommand{\cth}{\cos\theta_{\sss W}}
\newcommand{\sthq}{\sin^{2}\theta_{\sss W}}
\newcommand{\cthq}{\cos^{2}\theta_{\sss W}}

\newcommand{\proc}{\elp\elm $\rar t \bar{b} W^{\sss -}$}
\newcommand{\antiproc}{\elp\elm $\rar \bar{t} b W^{\sss +}$}
\newcommand{\proctt}{\elp\elm $\rar \bar{t} t$}
\newcommand{\procww}{\elp\elm $\rar W^{\sss +} W^{\sss -}$}

\begin{titlepage}
\pagestyle{empty}
\setcounter{page}{-1}

\begin{flushleft}
{\bf Preprint n.979} \hfill Dipartimento di Fisica \ \ \ \ \ \ \ \ \ \\
November 2, 1993 \hfill Universit\`a di Roma ``La Sapienza'' \\
hep-ph/9311263 \hfill I.N.F.N. - Sezione di Roma \ \ \ \ \ \  \\
Revised version: \\
December 29, 1993
\end{flushleft}

\vspace*{\fill}

\begin{center}
{\Large \bf Single Top Production at the Next Generation \\ Linear
\epem\ Colliders}

\vspace*{\fill}

\begin{tabular}[t]{c}
{\large S.~Ambrosanio\r{a,b,1}\ \ and \ B.~Mele\r{b,a,2}} \\
\\
{\r{a}\ \it Dipartimento di Fisica, Universit\`a ``La Sapienza'',} \\
{       \it P.le Aldo Moro 2, I-00185 Rome, Italy} \\
{\r{b}\ \it I.N.F.N., Sezione di Roma, Italy } \\
\end{tabular}
\end{center}

\vspace*{\fill}

\begin{abstract}
{\small \noindent
Present limits on the top mass from LEP1 and Tevatron point to a top quark
 that is considerably heavier than the $W$ vector boson
 in the standard model. Hence, \epem\ colliders with
$\rs \simeq 300$ GeV (the \cm\ energy foreseen at the first phase
of the Next Linear \epem\ Collider) could be well below the energy threshold
for real top-pair production.
We argue that, if this is the case, single top production through
the process \proc($\bar{t} b W^{\sss +}$), where $t\bar{b} (\bar{t}b)$
are produced mainly by means of a virtual $W$, becomes the dominant
top production mechanism.
Total cross sections and kinematical distributions are evaluated and numerical
results are given in ranges of $\mt$ and $\rs$  where single top
production can be of relevance. The relative importance of virtual-$W$
and virtual-$t$ contributions to the process is discussed.
}
\end{abstract}
\vspace*{\fill}

{\small
\r{1}\ {\it e-mail address: AMBROSANIO@VAXROM.ROMA1.INFN.IT}

\r{2}\ {\it e-mail address: MELE@VAXROM.ROMA1.INFN.IT}}
\end{titlepage}
\null
\pagestyle{empty}
\newpage

%
\pagestyle{myheadings}
\markboth{\it S.~Ambrosanio, B.~Mele / Single top production}{\it
 S.~Ambrosanio, B.~Mele / Single top production}

\section{Introduction} \label{sec:intro}

The Next \epem\ Linear Collider (NLC) will be an excellent place
where to study precision top quark physics. The \cm\ collision energy
range at this machine is foreseen to be between 300 GeV (most probably
in the first phase of operation) and 500 GeV, with a luminosity of
the order of  10\r{33}\ cm\r{-2}s\r{-1} \cite{hawaii,hamburg}.
Top production at NLC will be very interesting both at the threshold region,
where one expects some enhancement in the cross section due to the toponium
resonance remnants, and in the continuum region \cite{zerwas}.
 At threshold, one can measure $\mt$ with an accuracy better than
500 MeV. The shape of the excitation curve can be
precisely predicted in perturbative QCD. Also, it will be quite sensitive
to the perturbative strong coupling constant,  the top width and
 Higgs boson mass and Yukawa coupling to the top quark. On the other hand,
in the continuum top production well above the \tt\ threshold,
one will be able to test the
$t \to b W^{\sss +}$ coupling through the study of angular correlations
in the \tt\ decay products, as well as the existence of
anomalies in the top couplings and possible non-standard top decay modes.

Although as yet unobserved, the existence of the top quark
is strongly required by the theoretical
consistency of the standard model. Direct search
of top quark at Tevatron has yielded a lower limit $\mt > 113$ GeV
at 95\% C.L., for a top with standard-model decays \cite{top}.
The upgraded phase of the machine should either discover the top or
exclude the mass range up to 160-180 GeV.

On the other hand, precision electroweak measurements from LEP1,
$\mw$ determination at hadron colliders and neutrino scattering constrain
the top mass through  radiative corrections in the range
\beq
\mt = \left( 162^{+16+18}_{-17-21} \right) \; GeV
\label{eq:topmass}
\eeq
where the first error comes from experimental uncertainties and the second
corresponds to varying the Higgs mass between 60 GeV and 1 TeV, with a
central value of 300 GeV \cite{lefrancois}.

According to present limits on $\mt$, the top mass could well be in the range
(150-200) GeV. If this were the case, in the first phase of NLC
corresponding to a \cm\ energy of about 300 GeV, the collision energy could be
not sufficient for top-pair production.
In this situation, since the top is considerably heavier than
the $W$ vector boson, top quarks are mainly produced in association with
a $W$ and a $b$ quark through the processes
\beqn
\label{eq:procwtb}
e^{\sss +} e^{\sss -} & \to & t \bar{b} W^{\sss -}  \\
e^{\sss +} e^{\sss -} & \to & \bar{t} b W^{\sss +}.  \nonumber
\eeqn

 In fig.1, we show the Feynman diagrams corresponding to \proc\
 (charge conjugated graphs correspond to the reaction
 \antiproc). Aside the $tt^*$
channel that is responsible for top pair production above threshold,
in single top production also $bb^*$ and $WW^*$ channels, where a virtual $b$
quark and $W$ respectively are exchanged, contribute to the cross
section. In particular, the $WW^*$ turns out to be the dominant contribution
in almost the whole interesting $\rs$ range, that is
$\mt+\mw+\mb \le \rs \le 2 \mt$.
Of course, cross sections are ${\cal O}( \alpha_W^3 )$ and therefore
quite smaller than those relative to top pair production
(reported in Table 1) which are of the order of 1 pb at NLC
(\cf\ ref.\protect\cite{bern91}).

\begin{table}

\[ \begin{array}{|c||c|c|c|c|c|c|c|c|c|} \hline
{\bf \mt \: (GeV)} & {\bf 120} & {\bf 130} & {\bf 140} & {\bf 150} &
{\bf 160} & {\bf 170} & {\bf 180} & {\bf 190} & {\bf 200}  \\ \hline \hline
{\bf \rs = 300 \: GeV} & 1.37 & 1.16 & 0.85 & -   & --  & --  & --
			      & --  &   --   \\ \hline
{\bf 320} & 1.31 & 1.17 & 0.99 & 0.72 & -    & --   & --   & -- & -- \\ \hline
{\bf 340} & 1.23 & 1.13 & 1.01 & 0.85 & 0.62 & -    & --   & -- & -- \\ \hline
{\bf 360} & 1.14 & 1.07 & 0.98 & 0.88 & 0.74 & 0.54 & -    & -- & -- \\ \hline
{\bf 380} & 1.05 & 1.00 & 0.94 & 0.86 & 0.77 & 0.64 & 0.47 & -  & -- \\ \hline
{\bf 400} & 0.97 & 0.93 & 0.89 & 0.83 & 0.76 & 0.68 & 0.57 & 0.41 & - \\
\hline
{\bf 450} & 0.80 & 0.78 & 0.75 & 0.72 & 0.69 & 0.65 & 0.60 & 0.54 & 0.47 \\
\hline
{\bf 500} & 0.66 & 0.65 & 0.64 & 0.62 & 0.60 & 0.58 & 0.55 & 0.52 & 0.49 \\
\hline
\end{array} \]  \label{tab:tt}

\caption{\rm Total cross section (in pb) for the process
	 \protect\proctt\ at the Born level.}

\end{table}

In the present work we make a detailed study of single top production in
\epem\ collisions below the \tt\ threshold, namely when $\rs < 2 \mt$.
We mainly refer to the practical case of NLC with $\rs \simeq 300$ GeV
and an integrated luminosity of 10 fb\r{-1}, but also other possibilities
for the \epem\ collision energy  are considered. As far as the
top mass is concerned, we focus on the range
$150 \ltap \mt \ltap 200$ GeV,  with $\mt > \rs/2$.

We will neglect throughout our study toponium resonance effects near the \tt\
threshold.

As anticipated, cross sections are found to be rather moderate
(of the order of a few fb's) and marginal
for the study of top properties at the foreseen NLC luminosity.
Nevertheless, single top production  could be a relevant background for rare
processes and possible new physics signal at future \epem\ colliders.
In fact, as a result of the fast top decay, one observes a
$W^{\sss +}W^{\sss -}\bar{b}b$ final state, that is
a  possible background for any process involving multi-vector boson
production. For instance, standard-model processes like
\mbox{\epem\ $\to WWZ$} \cite{wwz} and \mbox{\epem\ $\to WWH$} \cite{mele93}
involve the same final state.
Also, analogous signals could correspond to new physics effects, \eg\
chargino production in supersymmetric models \cite{majerotto}.

Single top production in \epem\ collisions through the process \epem\ $\to
t \bar{b} e^{\sss -} \bar{\nu}_{\sss e}$ was considered in ref.\cite{katuya}
for $\mt = (30$-60) GeV at PETRA and TRISTAN energies. With this
kinematical configuration,
a real $W$  in the final state (as in \proc) is not allowed and
single top production occurs mainly through the higher-order reaction
$\gamma$\elp\ $\to t \bar{b} \bar{\nu}_{\sss e}$, where an
almost-real photon is
radiated by the initial \elm\ beam.
Therefore, the corresponding production rates are quite small
(and undetectable).
For instance, at $\rs = 50$ and 65 GeV total cross sections of
$1.1 \times 10^{-5}$ fb  and $2.9 \times 10^{-3}$ fb, respectively,
are found for $\mt = 40$ GeV.

After completion of our work, we learnt of two recent studies
concerning single top production at LEP200 \cite{panella,raidal}.
In ref.\cite{panella}, analytical expressions for cross sections of
the reaction \epem\ $\to t \bar{b} e^{\sss -} \bar{\nu}_{\sss e}$ are given
for the photon initiated process, as well as for the $WW^*$ and $tt^*$
contributions (where $W\to e \nu$).
Unfortunately, below the \tt\ threshold, event rates
are found to be too small for $\mt$ larger than the present Tevatron limit,
even for \cm\ energies of 220-250 GeV
(that are, however, well above the foreseen $\rs$ at LEP200).
In this work, as in ref.\cite{katuya}, the authors concentrate on $e\nu$
final states, that replace the real $W$ in the reaction \proc. This is because
with the LEP200 \cm\ energy, for most of the interesting $\mt$ range, one
can not afford to produce a top quark plus a {\it real} $W$. Hence,
the higher-order $\gamma$\elp\ $\to t \bar{b} \bar{\nu}_{\sss e}$
process becomes the dominant channel.
Instead, in the case of \proc\ at higher \cm\ energies, that we consider
here, the final $e\nu$ state can be replaced by other leptonic and hadronic
decays of the $W$, that are equally interesting and increase the
production rate.
As far as $tt^*$ and $WW^*$ channels are concerned, we compared the output of
our programs with the results of ref.\cite{panella} and found complete
agreement at LEP200 energies.

In ref.\cite{raidal}, a similar study for single top production at LEP200
has been carried out with drastically different results with respect to
ref.\cite{panella}. We think that
such disagreement is due to a mis-treatment of the photon collinear
singularity in the process \epem\ $\to t \bar{b} e^{\sss -} \bar{\nu}_{\sss e}$
that largely overestimates the contribution from diagrams with a photon
exchange in the t-channel.

We stress that the NLC regime, that we are considering here, will provide
for the first time the environment to study single top production
{\it at the leading order} and with observable rates.
Furthermore, contrary to the previous studies, we include the $bb^*$
contribution and study in detail kinematical distributions.

After a short description of our computation  method, in section 2,
we will discuss the behaviour of total cross sections for the reaction
\proc$(\bar{t} b W^{\sss +})$ versus $\rs$ and $\mt$, stressing the relative
importance of various contributions to the matrix element squared.
In section 3, we will study the distribution in the $Wb$
invariant mass and the angular distribution of the top with respect to the
beam. Once more, contributions from different diagrams will be emphasized.
In section 4, we will give our conclusions.

\section{Total cross sections}
\label{sec:totcross}

In fig.1, the seven Feynman diagrams contributing to the process \proc\
at tree level are shown.
As a result, one has 28 terms, including interferences, in the expression
for the matrix element squared.

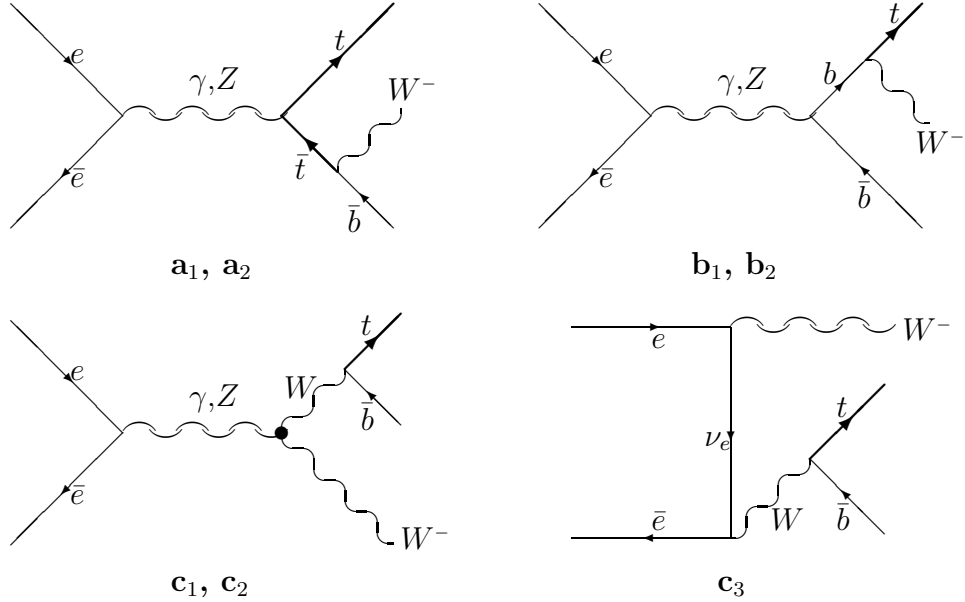
\begin{figure}

\begin{picture}(45000,22000)

\drawline\photon[\E\REG](8000,18000)[6]
\drawline\fermion[\NW\REG](\photonfrontx,\photonfronty)[6000]
\drawarrow[\SE\ATBASE](\pmidx,\pmidy)
\drawline\fermion[\SW\REG](\photonfrontx,\photonfronty)[6000]
\drawarrow[\SW\ATBASE](\pmidx,\pmidy)

\thicklines 

\drawline\fermion[\NE\REG](\photonbackx,\photonbacky)[6000]
\drawarrow[\NE\ATBASE](\pmidx,\pmidy)
\drawline\fermion[\SE\REG](\photonbackx,\photonbacky)[3000]
\drawarrow[\NW\ATBASE](\pmidx,\pmidy)

\thinlines

\drawline\photon[\NE\FLAT](\pbackx,\pbacky)[4]
\drawline\fermion[\SE\REG](\photonfrontx,\photonfronty)[3000]
\drawarrow[\NW\ATBASE](\pmidx,\pmidy)

  \put(6000,20000){$e$}
  \put(6000,15300){$\bar{e}$}
  \put(10500,19000){$\gamma$,$Z$}
  \put(16000,20500){$t$}
  \put(14500,15600){$\bar{t}$}
  \put(16500,13600){$\bar{b}$}
  \put(18000,18600){\Wm}

  \put(9800,12000){\bf a$_1$, a$_2$}

\drawline\photon[\E\REG](28000,18000)[6]
\drawline\fermion[\NW\REG](\photonfrontx,\photonfronty)[6000]
\drawarrow[\SE\ATBASE](\pmidx,\pmidy)
\drawline\fermion[\SW\REG](\photonfrontx,\photonfronty)[6000]
\drawarrow[\SW\ATBASE](\pmidx,\pmidy)
\drawline\fermion[\SE\REG](\photonbackx,\photonbacky)[6000]
\drawarrow[\NW\ATBASE](\pmidx,\pmidy)
\drawline\fermion[\NE\REG](\photonbackx,\photonbacky)[3000]
\drawarrow[\NE\ATBASE](\pmidx,\pmidy)
\drawline\photon[\SE\FLAT](\pbackx,\pbacky)[4]
\thicklines
\drawline\fermion[\NE\REG](\photonfrontx,\photonfronty)[3000]
\drawarrow[\NE\ATBASE](\pmidx,\pmidy)
\thinlines

\put(26000,20000){$e$}
\put(26000,15300){$\bar{e}$}
\put(30500,19000){$\gamma$,$Z$}
\put(34500,19200){$b$}
\put(37000,21500){$t$}
\put(35800,14500){$\bar{b}$}
\put(38000,16600){\Wm}

\put(29500,12000){\bf b$_1$, b$_2$}

\drawline\photon[\E\REG](8000,6000)[6]
\drawline\fermion[\NW\REG](\photonfrontx,\photonfronty)[6000]
\drawarrow[\SE\ATBASE](\pmidx,\pmidy)
\drawline\fermion[\SW\REG](\photonfrontx,\photonfronty)[6000]
\drawarrow[\SW\ATBASE](\pmidx,\pmidy)
\drawline\photon[\NE\FLAT](\photonbackx,\photonbacky)[4]
\thicklines
\drawline\fermion[\NE\REG](\photonbackx,\photonbacky)[3000]
\drawarrow[\NE\ATBASE](\pmidx,\pmidy)
\thinlines
\drawline\fermion[\SE\REG](\photonbackx,\photonbacky)[3000]
\drawarrow[\NW\ATBASE](\pmidx,\pmidy)
\drawline\photon[\SE\FLIPPEDFLAT](\photonfrontx,\photonfronty)[7]

  \put(14000,6000){\circle*{500}}

\put(6000,8000){$e$}
\put(6000,3300){$\bar{e}$}
\put(10500,7000){$\gamma$,$Z$}
\put(14100,7300){$W$}
\put(17000,9700){$t$}
\put(17000,6100){$\bar{b}$}
\put(18500,1600){\Wm}

\put(9800,0){\bf c$_1$, c$_2$}

\drawline\fermion[\E\REG](25000,10000)[6000]
\drawarrow[\E\ATBASE](\pmidx,\pmidy)
\drawline\photon[\E\REG](\pbackx,\pbacky)[6]
\drawline\fermion[\S\REG](\photonfrontx,\photonfronty)[8000]
\drawarrow[\S\ATBASE](\pmidx,\pmidy)
\drawline\photon[\NE\FLIPPEDFLAT](\pbackx,\pbacky)[5]
\thicklines
\drawline\fermion[\NE\REG](\photonbackx,\photonbacky)[4000]
\drawarrow[\NE\ATBASE](\pmidx,\pmidy)
\thinlines
\drawline\fermion[\SE\REG](\photonbackx,\photonbacky)[4000]
\drawarrow[\NW\ATBASE](\pmidx,\pmidy)
\drawline\fermion[\W\REG](\photonfrontx,\photonfronty)[6000]
\drawarrow[\W\ATBASE](\pmidx,\pmidy)

\put(28000,9200){$e$}
\put(28000,2300){$\bar{e}$}
\put(30000,5500){\ve}
\put(37500,9500){\Wm}
\put(35000,6600){$t$}
\put(35000,2300){$\bar{b}$}
\put(32500,2300){$W$}

\put(30500,0){\bf c$_3$}

\end{picture}

\caption{\rm Feynman diagrams for the process \protect\proc.
           a$_i$: diagrams with top exchange, b$_i$: diagrams with
	   $b$-quark exchange, c$_i$: diagrams with $W$ exchange.
           $i=1$: $\gamma$ exchange in the s-channel,
	   $i=2$: \Z\ exchange in the s-channel,
           $i=3$: \ve\ exchange in the t-channel.}

\end{figure}
 Aside the graphs {\bf a}$_{1,2}$, that correspond to the
relevant diagrams for production of a pair of real top quarks, here we have
$b$- and $W$-exchange diagrams, where the top quark comes from the decay
of a virtual $b$ quark and $W$ vector boson respectively. In the latter
case, aside the $\gamma$ and $Z$ contributions, we have a further
graph coming from $W$-pair production through neutrino exchange in the
t-channel.
 Note that the virtual $W$ and, to a greater extent, $b$ momenta in
{\bf b$_i$}- and \mbox{{\bf c$_i$}-type} diagrams are always far from
their mass shell, since an heavy top has to be produced.
As for the \mbox{{\bf a$_i$}-type} graphs, if the \cm\
 energy is above the \tt\ threshold, they reproduce real \tt\
production with subsequent $\bar{t} \to \bar{b}W^{\sss -}$ decay on the
$\bar{t}$ mass shell. Hence, the corresponding contribution turns out to be
quite enhanced.
On the other hand, when $\rs$ is well below the \tt\ threshold, the
contributions from virtual-$W$ (and -$b$) exchange could be, in a first
qualitative approach, not negligible with respect to the $t$ exchange.
Indeed, we will show that $W$ exchange is dominant for most of the
$\rs$ range of interest for single top production.

 Numerical results for total cross sections and distributions for the process
\proc\ have been obtained in the following way.
We used the program \mbox{{\sc CompHEP} 2.3-5} \cite{comphep}
in order to get the  matrix element  squared for \proc\
at the Born level, in a symbolic form and in a FORTRAN compatible
form as well.
Then we developed some software (based on VEGAS \cite{vegas}),
 performing a numerical Monte Carlo phase space (\ps) integration
over the three-massive-particle final state with three different masses.
We chose the VEGAS input parameters in such a way
as to obtain a three-digit accuracy on the \ps\ integration,
 that is \mbox{3-dimensional} for distributions and
 \mbox{4-dimensional} for total cross sections.
 Firstly, we performed the \ps\ integration
on the solid angles of the $Wb$ system in its \cm\ frame.
The relevant formula for the invariant distributions for top variables is then
\beq
D_{top} = E_{\sss t} \frac{d^{3}\sigma}{d{\bf p}_{\sss t}} =
\frac{\beta_{\sss W}}{(4\pi)^{5}s}
\left(1 + \frac{\mw^2 - \mb^2}{\mwb^2} \right)
\int \overline{ | {\cal M} |^2} \: d\!\cos\theta_{\sss W} \: d\phi_{\sss W}
\label{eq:d3top}
\eeq
where
\beq
\beta_{W} =
\sqrt{1- 4 \frac{\mwb^2 \mw^2}{(\mwb^2 + \mw^2 -\mb^2)^2}}.
\eeq
 $\mwb$ is the
invariant mass of the $Wb$ system, which is linked to the top energy,
$E_{\sss t}$, in the \epem\ \cm\ system by
\beq
\mwb = \sqrt{ s + \mt^2 - 2 \rs E_{\sss t}},
\label{eq:topvar}
\eeq
and $\theta_{\sss W}$, $\phi_{\sss W}$ are
the polar and azimuthal
angle of the $W$ with respect to the \elm\ beam direction in the $Wb$ rest
frame. In eq.(\ref{eq:d3top}), the squared matrix element
$\overline{| {\cal M} |^2}$
(averaged over initial and final particle polarizations) for the process \proc\
has been expressed in terms of $\theta_{\sss W}$, $\phi_{\sss W}$,
$\mwb$ and $\costop$, where $\theta_{\sss t}$ is the
angle between the top 3-momentum and the \elm\ one, in the \elp\elm\ \cm\
frame. Then, one has
\beq
\frac{d^{2}\sigma}{d\mwb \: d\!\costop} = 2 \pi |{\bf p}_{\sss t}|
\frac{\mwb}{\rs} D_{top}
\label{eq:d2top}
\eeq
At this point, total cross sections are obtained by
performing the remaining two integrations over $\mwb$ and $\costop$.
We checked the stability of the numerical integration by choosing different
 sets of independent kinematical variables.
We found that the above  choice is a very good one in order to optimize
the estimated relative error on total cross section.
Furthermore, we found that the choice of the integration variable $\mwb$
has a crucial role in
reproducing cross sections  above the \tt\ pair threshold, starting from
single top production formulae. Indeed, above the \tt\ threshold the bulk
of the cross section comes from the pole
corresponding to  $\mwb \simeq \mt$ in the  $t$ propagator in
diagrams {\bf a$_{1,2}$}.

In our computation, we used the following set of parameter values
\[ \begin{array}{rcllrcll}
\mz   & = & 91.18 & GeV, & \mw 	& = & 80.1 & GeV, 	\\
\wz   & = & 2.49  & GeV, & \ww 	& = & 2.12 & GeV, 		\\
\mb   & = & 5.00  & GeV, & \alpha_{\sss EM} & = & 1/128	&  , 	\\
\sthq & = & 0.232 &  ,   & V_{\sss tb} & = & 0.999 &  .
\end{array} \]
The precise value of the top width $\wt$, which enters
the amplitudes relative to the graphs {\bf a$_{1,2}$}
through the top propagator \mbox{$1/(p^{2} - \mt^{2} + i \mt \wt$)},
is not important when one considers  \proc\ well below
the \tt\ threshold.
On the contrary, if one extends the \proc\ formulae to values of
$\rs$ around and above the \tt\ threshold, (\ie\ integrates over the
pole in the top propagator), the results for cross section are
rather sensitive to the value of $\wt$. In order to recover through
our computation also the Born cross sections for real \tt\
production at $\rs \gtap 2\mt$, we have approximated the top
total width  with the tree level width for $t \to Wb$
 \footnote{Note that $\Gamma(t \to b W) \simeq 0.998 \:
\protect\Gamma_{\sss t}^{\sss TOT}$ irrespective of $\mt$
(for $\mt \gtap 120$ GeV)
and that the inclusion of QCD, electroweak radiative corrections and finite
$W$-width effects decreases $\wt$ by
about 8-9\%  \cite{kuhn}.} \cite{kuhn81}:
\beq
\Gamma_{\sss t} \simeq \Gamma(t \to b W) \simeq
\frac{G_{F} \mt^3}{8\pi \sqrt{2}} \left( 1 - \frac{\mw^2}{\mt^2} \right)^2
\left( 1 + 2 \frac{\mw^2}{\mt^2} \right) .
\label{eq:topwidth}
\eeq

 In this way, starting from our formulae for \proc, to a very
good approximation, we obtain for $\rs \ge 2 \mt$
\beq
\sigma(Wtb) = \sigma(\bar{t}t) \cdot BR(t \to Wb)
\simeq \sigma(\bar{t}t),
\label{eq:wtbvstt}
\eeq
where $\sigma(\bar{t}t)$ is the tree level cross section for \proctt\
reported in Table 1.

Values of $\wt$ obtained through
eq.(\ref{eq:topwidth}) are shown versus $\mt$ in Table 2.

\begin{table}

\[ \begin{array}{|c||c|c|c|c|c|c|c|c|c|} \hline
{\bf \mt \: (GeV)}          & {\bf 120} & {\bf 130} & {\bf 140}
& {\bf 150} & {\bf 160} & {\bf 170} & {\bf 180} & {\bf 190} & {\bf 200}
\\ \hline \hline
\Gamma(t \to b W) \: (GeV)  & 0.327    & 0.485       & 0.671
& 0.885     & 1.13      & 1.41     & 1.71     & 2.06        & 2.44
\\ \hline
\end{array} \] \label{tab:topwidth}

\caption{\rm Top quark width in the Born approximation for several values of
	$\mt$.}

\end{table}

In our computation, we will not take into account  resonance effects
near \tt\ threshold and, as a consequence, our results can be
trusted only when $2\mt-\rs \gg \wt$ below threshold.
In practice, our  results  should be rather accurate at tree level
in the region $\rs \ltap 2 \mt- 3 \wt$ and $\rs \gtap 2 \mt$.

 Total cross sections for \proc\ are presented in Table 3 and in figs.2-4.
In Table 3 we report the numerical results of our programs
(in fb), summing up over both
\protect\proc\ and $\protect\bar{t} b \protect W^{\sss +}$
contributions.
We restrict ourselves to the range 150 $\leq \mt \leq 200$ GeV
and  $240 \leq \rs \leq 400$ GeV, thus including in our study the
possibility of intermediate \epem\ collision energies between LEP200
and NLC at $\rs = 300$ GeV.

\begin{table}
\label{tab:wtb}
\[ \begin{array}{|c||c|c|c|c|c|c|} \hline
{\bf \mt \: (GeV)} & {\bf 150} & {\bf 160} & {\bf 170} & {\bf 180}
	  & {\bf 190} & {\bf 200}  \\ \hline \hline
{\bf \rs = 240 \: GeV}
		& 0.018 &  --   &   --  &  --   &  --   &  --   \\ \hline
{\bf  260}  	& 0.89  & 0.19  & 0.010 &  --   &  --   &  --   \\ \hline
{\bf  280}  	& 4.2   & 1.5   & 0.50  & 0.11  & 0.006 &  --   \\ \hline
{\bf  300}  	& \ttb  & 5.2   & 2.2   & 0.90  & 0.31  & 0.070 \\ \hline
{\bf  320}  	& \ttb  & \ttb  & 6.3   & 2.8   & 1.3   & 0.59  \\ \hline
{\bf  340}  	& \ttb  & \ttb  & \ttb  & 7.2   & 3.4   & 1.8   \\ \hline
{\bf  360}  	& \ttb  & \ttb  & \ttb  & \ttb  & 8.1   & 4.0   \\ \hline
{\bf  380}  	& \ttb  & \ttb  & \ttb  & \ttb  & \ttb  & 9.0   \\ \hline
{\bf  400}   	& \ttb  & \ttb  & \ttb  & \ttb  & \ttb  & \ttb  \\ \hline
\end{array} \]

\caption{\rm Total cross section (in fb) for single top production.
         Contributions from both  \protect\proc\  and
  	 $\protect\bar{t} b \protect W^{\sss +}$ processes are considered.
	 Entries in the range $\protect\rs \geq 2
         \protect\mt$, where top-pair production
         is dominant, are marked by ``\protect\tt''.}

\end{table}

In Table 3, results are shown only for
 $\mt +\mw +\mb < \rs < 2 \mt$, that is where single top production is the
dominant top production mechanism.
 Here, cross sections turn out to be about two orders of
magnitude smaller than those for top-pair production
above threshold (\cf\ Table 1). In particular,
by varying $\mt$ from 150 GeV up to 200 GeV, one
gets cross sections from 4 to 9 fb for $\rs = 2 \mt - 20$ GeV
and from 0.9 to 4 fb for $\rs = 2 \mt - 40$ GeV.
For 180 GeV $< \mt <$ 200 GeV,  we still have cross sections of the
order of 1 fb or more for $\rs = 2 \mt - 60$ GeV.
 Assuming an integrated luminosity of 10 fb\r{-1}, this corresponds
to the production of 10-100 single top events. Hence, even after inclusion of
experimental cuts and efficiencies, one should get observable
single top event rates.

\begin{figure}

\setlength{\unitlength}{1truecm}

\begin{picture}(12.0,18.0)
 \put(0.3,4.5){\special{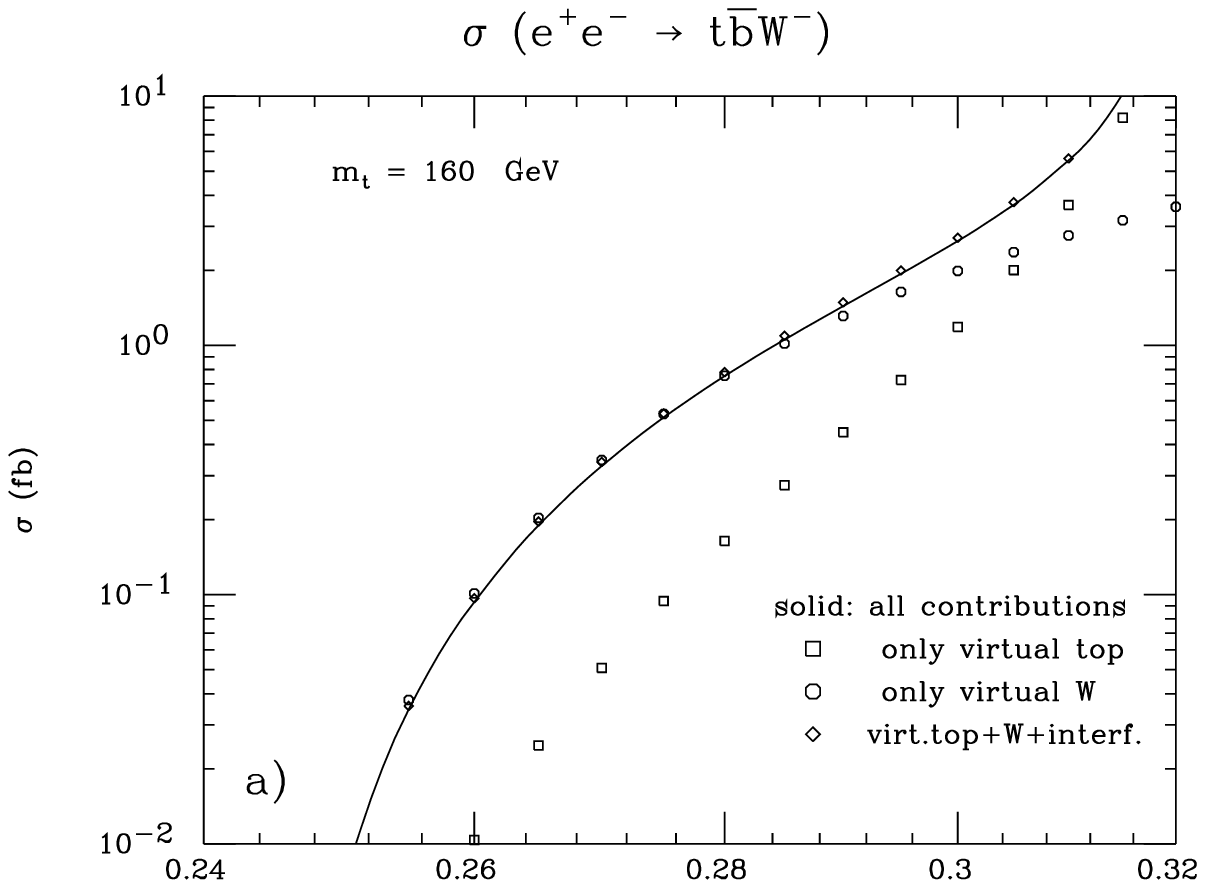}}
 \put(0.3,-4.5){\special{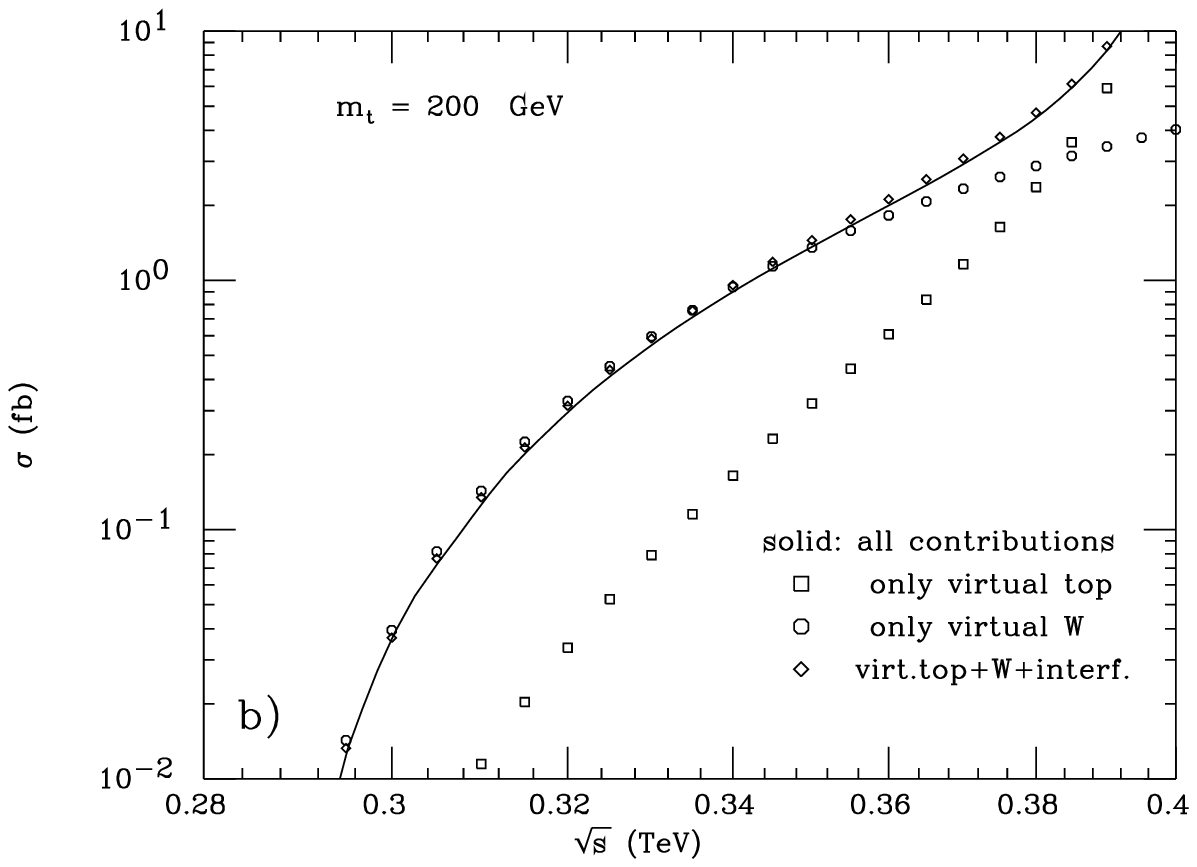}}
\end{picture}

\caption{\rm Various contributions to the total cross section of for
    	 \protect\proc\ when $\protect\mt+\protect\mw+\protect\mb \le
	 \protect\rs \le 2\mt$. a) $\protect\mt = 160$ GeV;
	 b) $\protect\mt = 200$ GeV. \protect\\ Squares
	 and circles represent respectively contributions from
	 diagrams with virtual-top and virtual $W$-exchange. Thunders
	 account for both the previous two (plus interferences).
         The solid line shows total cross section including all
         contributions in fig.1.}

\protect\label{fig:wtb2}

\end{figure}
 In fig.\ref{fig:wtb2}, the behaviour of various contributions to the cross
section versus $\rs$, in the region below the \tt\ threshold, is shown
for $\mt=160$ GeV (fig.\ref{fig:wtb2}a) and $\mt = 200$ GeV
(fig.\ref{fig:wtb2}b). Squares represent the contribution to the total cross
section coming from diagrams  {\bf a$_{1,2}$} in fig.1, in which a virtual top
is exchanged, while the
circles stand for the virtual-$W$ contribution, namely the one coming from
 diagrams {\bf c$_{1,2,3}$}. Thunders account for virtual-top and
virtual-$W$ diagrams (including interferences between the two).
Finally, the solid line refers to the complete computation that takes into
account all diagrams in fig.1.

 Fig.\ref{fig:wtb2} shows very clearly that the contribution of
virtual-$W$ diagrams is very important. Indeed, it is larger than
the one from virtual-top diagrams for  $\rs \ltap 2 \mt - (7 \div 10)\wt$.
The latter can be even neglected for still lower $\rs$, where it is
about an order of magnitude lower than the virtual-$W$ contribution.
On the other hand, the contribution coming from  diagrams {\bf b$_{1,2}$}
with  virtual-$b$ exchange is very small everywhere. For this reason,
in our plots the thunders are well  superimposed to the solid line.
This is due to  the $b$-quark propagator suppression of the amplitudes
relative to diagrams {\bf b$_{1,2}$},  since the $b^*$ has to decay
in a heavy $tW$ final state. Concerning interferences between virtual-top
and virtual-$W$ diagrams, they are destructive and in general
small. This is no more true in the $\rs$ region  below the \tt\ threshold
by few $\wt$'s, where virtual-top and virtual-$W$ contributions becomes both
sizable and comparable.

\begin{figure}

\setlength{\unitlength}{1truecm}

\begin{picture}(12.0,9.0)
 \put(0.3,-4.5){\special{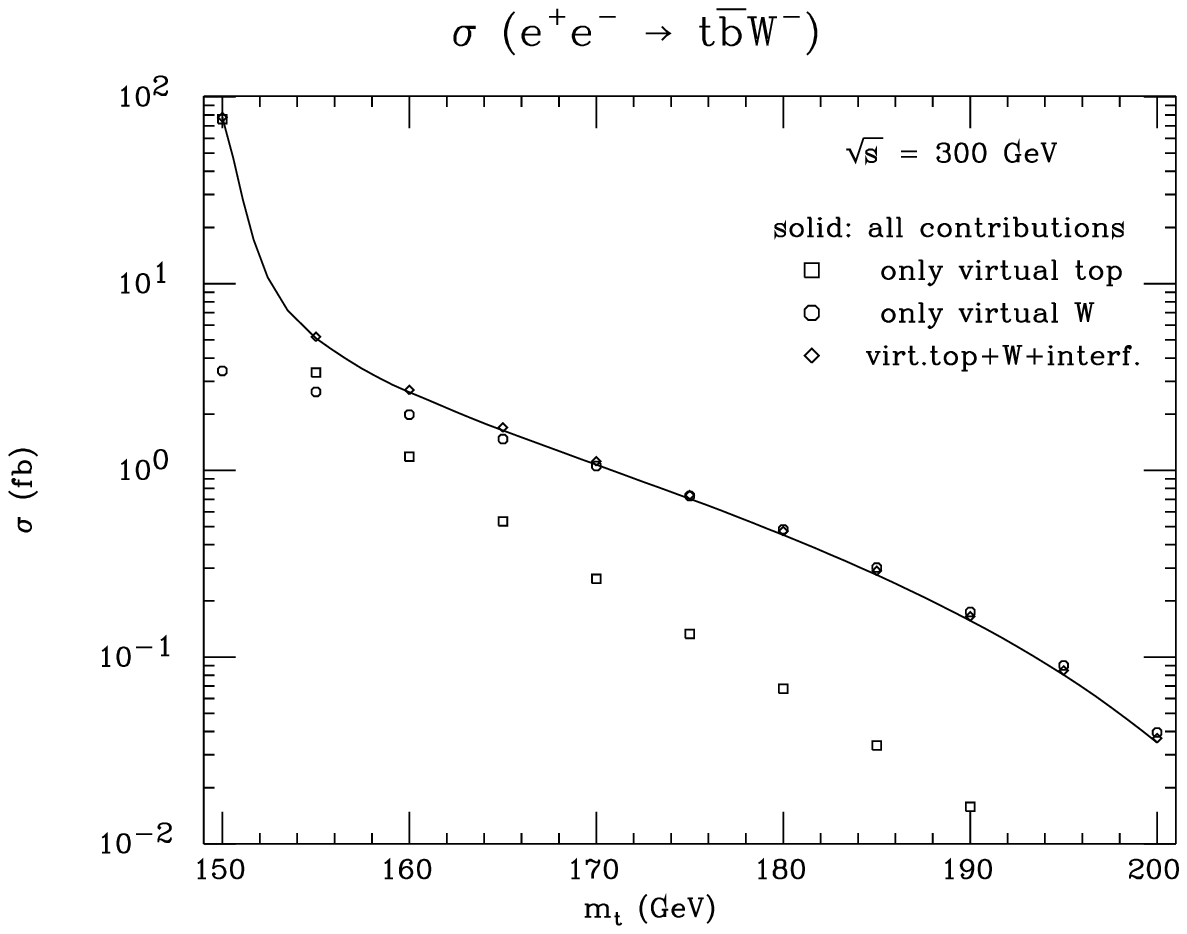}}
\end{picture}

\caption{\rm Various contributions to the total cross section of the process
	 \proc\ for $\protect\rs = 300$ GeV. Notations are the
	 same as in fig.\protect\ref{fig:wtb2}.}

\protect\label{fig:wtb3}

\end{figure}
 In fig.\ref{fig:wtb3}, the various contributions to total cross
section are analyzed as functions of $\mt$, for fixed $\rs = 300$ GeV,
that could be of interest for the first phase of NLC operation.
One can see that, when $\mt \gtap 165$ GeV, considering only the
virtual-$W$ contribution is a very good approximation for
the total cross section. In particular, for $165 \ltap \mt \ltap 190$ GeV
the usual $tt^*$ contribution would underestimate the production rates
by a factor from 3 to 10.
 After the \tt\ threshold region, where the
virtual-top contribution dominates, we have an intermediate $\mt$
range, $155 \ltap \mt \ltap 160$ GeV, where the virtual-top and
virtual-$W$ contributions are separately comparable and have
 destructive interference in the total.

Of course, event rates decrease  when $\mt$ grows.
At $\rs=300$ GeV, with an integrated luminosity of 10 fb\r{-1}
and  summing over $t$ and $\bar{t}$ events, one can hope to detect
through single top production a top quark with $\mt$ up to about 185 GeV.

\begin{figure}

\setlength{\unitlength}{1truecm}

\begin{picture}(12.0,9.0)
 \put(0.3,-4.5){\special{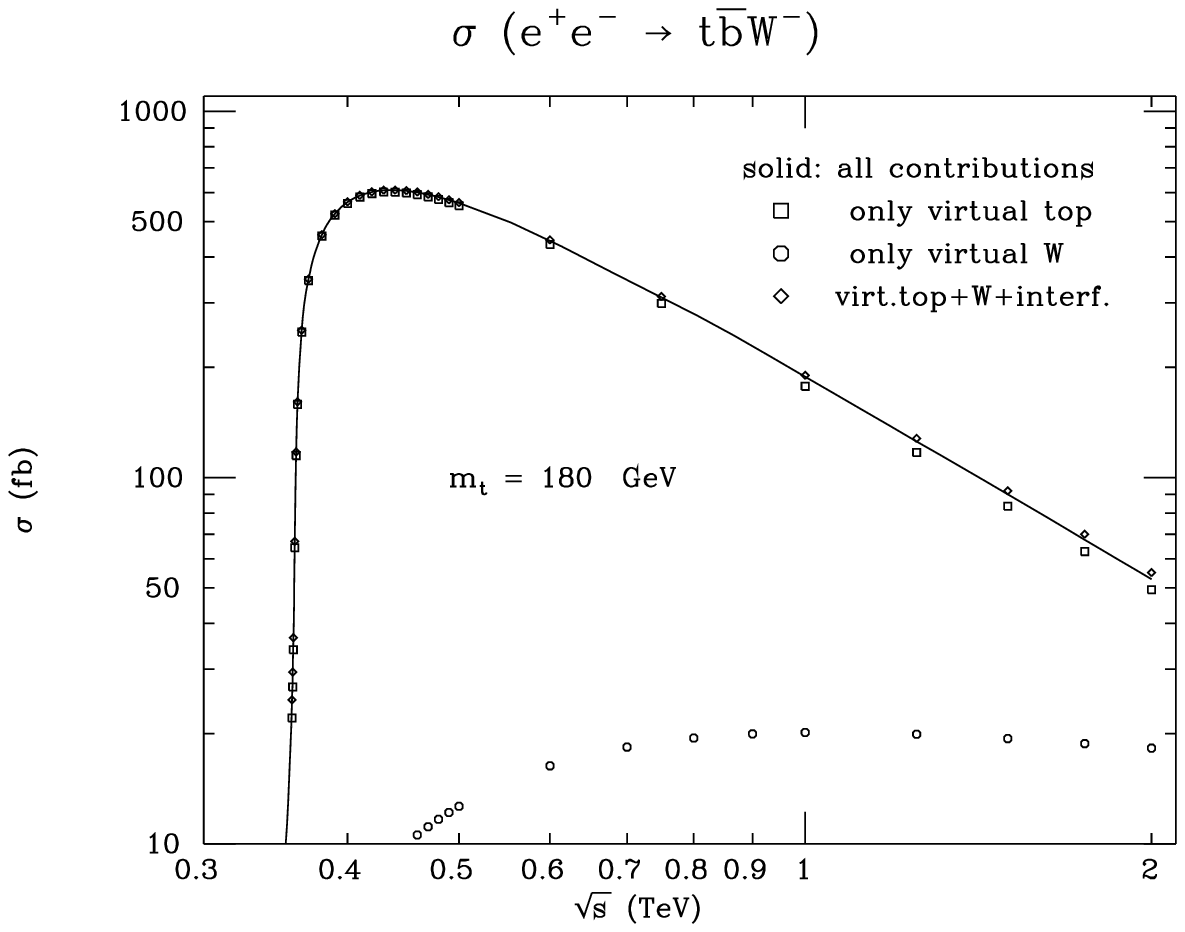}}
\end{picture}

\caption{\rm Various contributions to the total cross section of the process
	 \proc\ for $\protect\mt = 180$ GeV and
	 $\protect\rs \ge \protect 2\protect\mt$. Notations are
	 as in fig.\protect\ref{fig:wtb2}.}
\protect\label{fig:wtb4}

\end{figure}
 In fig.\ref{fig:wtb4}, we plot the \proc\ cross section versus $\rs$
above the \tt\ threshold and up to $\rs= 2$ TeV, for $\mt = 180$ GeV.
Of course, $\sigma(\bar{t}t)$ is recovered in this case, since we assumed
 BR$(t \to Wb)\simeq 1$. Note that,
although in general negligible in this case,
the relative importance of single top production through virtual-$W$
increases at high energies.
Indeed, as the contribution of {\bf a$_{1,2}$} diagrams in fig.1
quickly decreases after crossing the \tt\ threshold peak,
the contribution of {\bf c$_{1,2,3}$} diagrams keeps  growing up to
$\rs \simeq 1$ TeV and then starts to decrease rather slowly.
This is due mainly to the presence of trilinear boson vertex
in the diagram {\bf c$_3$}.
As a result, \eg\ we have that at $\rs = 2$ TeV the cross section for
single top production through a virtual $W$ only would be about one third of
the continuum \tt\ cross section.
In reality, in this
energy region there is a large interference effect between $WW^*$
and $tt^*$ diagrams that restores almost completely the usual
\tt\  cross section coming from the production of two real top quarks (\cf\
fig.\ref{fig:wtb4}).

\section{Kinematical distributions}
\label{sec:distr}
\begin{figure}

\setlength{\unitlength}{1truecm}

\begin{picture}(12.0,12.5)
 \put(-0.5,-3.5){\special{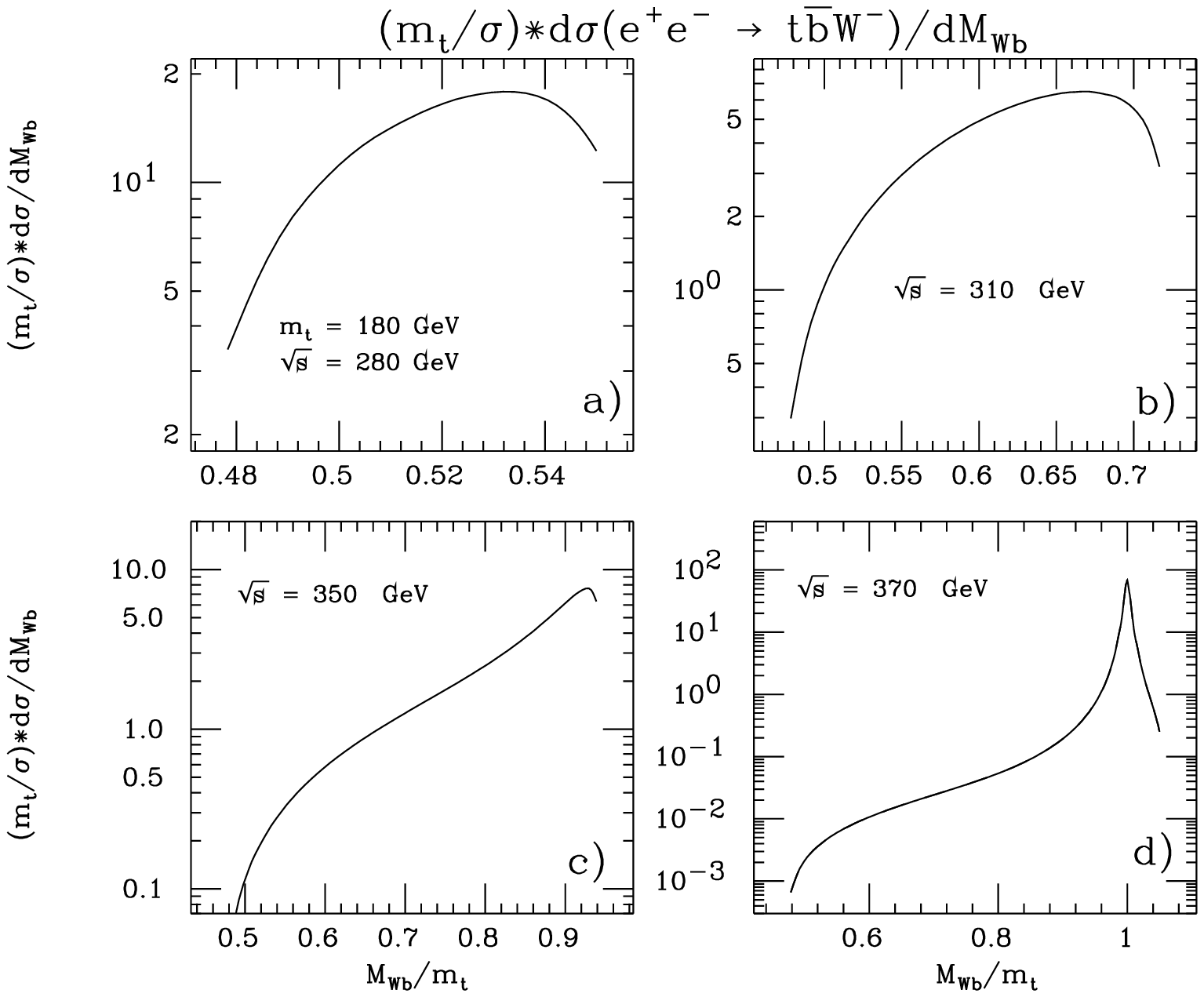}}
\end{picture}

\caption{\rm Distribution in the invariant mass of the $Wb$  system
    $\protect\mwb$ for \proc\ when $\mt=180$ GeV and:
	 a) $\protect\rs=280$ GeV; b) $\protect\rs=310$ GeV;
	 c) $\protect\rs=350$ GeV; d) $\protect\rs=370$ GeV.}

\protect\label{fig:wtb5}

\end{figure}
 In this section, we study kinematical distributions that characterize
the final state in single top production. Particular emphasis is given
to the comparison with analogous distributions above the \tt\ threshold.
In fig.\ref{fig:wtb5}, distributions in the
$Wb$ invariant mass $\mwb$  for the process \proc\ are shown for
$\mt =180$ GeV and different
values of $\rs$, below or just above the \tt\ threshold  ($\rs = 280, \;
310, \; 350, \; 370$ GeV in figs.\ref{fig:wtb5}a,b,c,d respectively).
 The curves are normalized in such a way  as to be adimensional
and to have an integral equal to 1 when integrated over the whole $\mwb/\mt$
range, that is
\beq
\frac{\mw + \mb}{\mt} \le \frac{\mwb}{\mt} \le \frac{\rs}{\mt} -1.
\eeq
Well under the \tt\ threshold, the $\mwb$ distribution is rather broad
and has a maximum
at values much lower than $\mt$. This is due to the dominance in this region
of virtual-$W$ contributions, for which the invariant mass of the $Wb$ pair
is not correlated to $\mt$ or to any other resonance.
Then, as the \cm\ energy approaches the \tt\ threshold,
one has the progressive appearance of the $\bar{t}$-resonance peak
at $\mwb \simeq \mt$, with increasingly small tails at lower $\mwb$.
This, of course, is due to the $t^{*}$ propagator in the virtual-top diagrams,
that dominate onto other contributions near and above the \tt\
threshold, where most of the cross section comes from the production
of a real top and consequent decay in $Wb$.

\begin{figure}

\setlength{\unitlength}{1truecm}

\begin{picture}(12.0,18.0)
 \put(0.3,4.5){\special{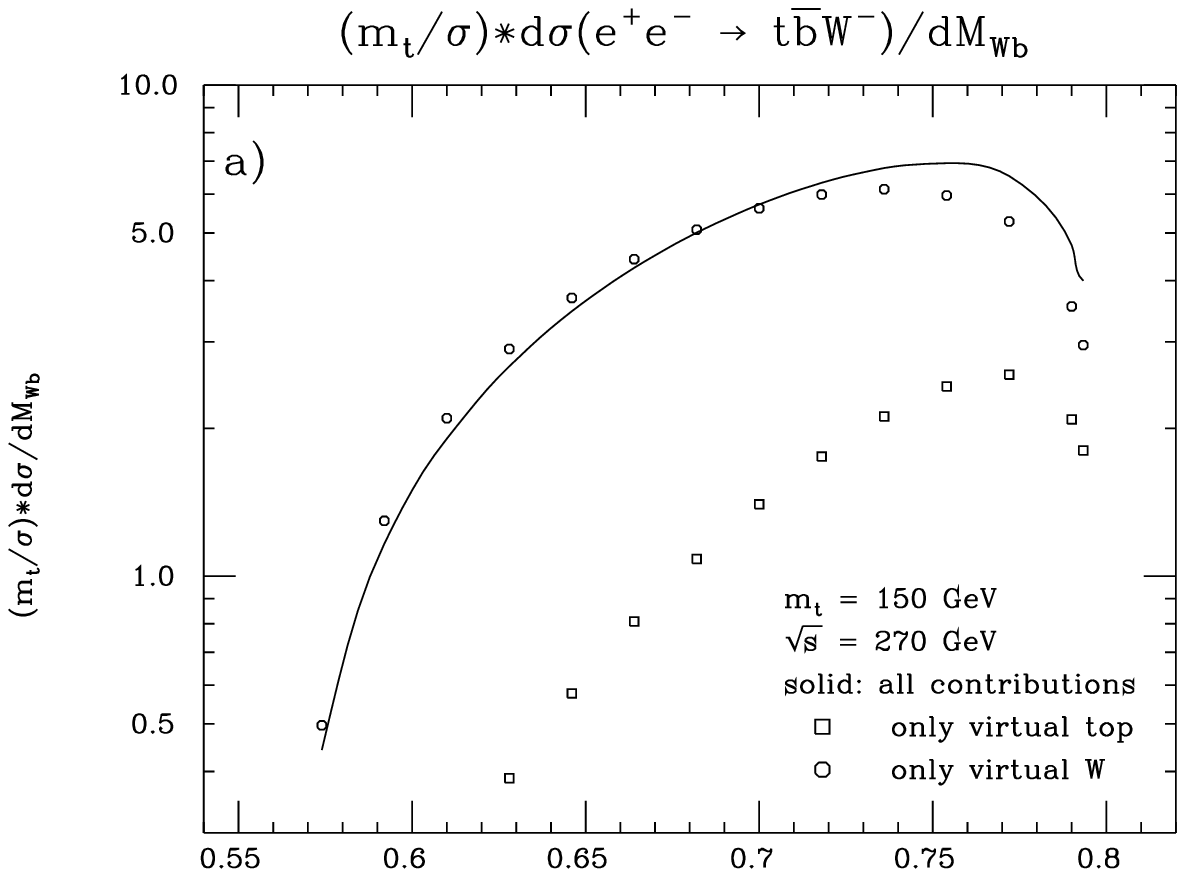}}
 \put(0.3,-4.5){\special{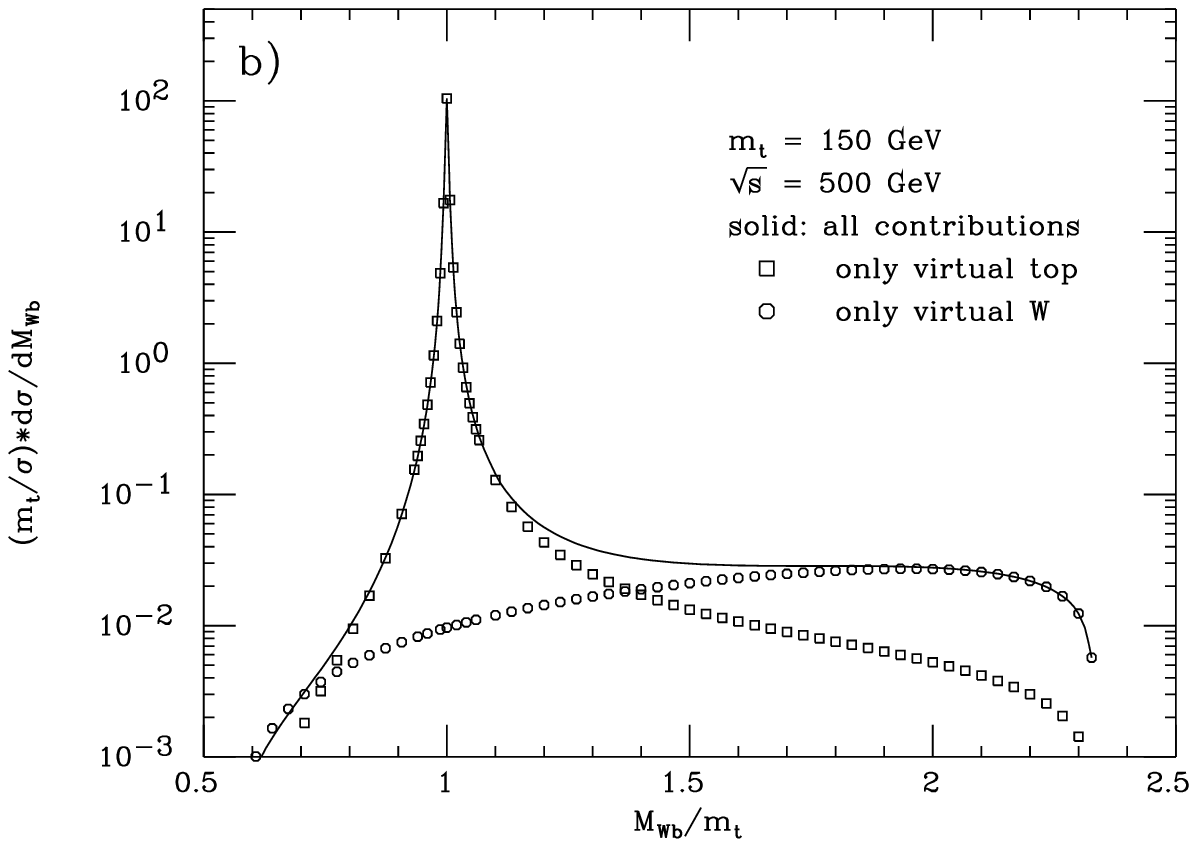}}
\end{picture}

\caption{\rm Various contributions to the distribution for the invariant mass
	 $\protect\mwb$ in the case $\mt = 150$ GeV. a) $\protect\rs =
	 270$ GeV, b) $\protect\rs = 500$ GeV. Notations are as in
	 fig.\protect\ref{fig:wtb2}.}

\protect\label{fig:wtb6}

\end{figure}

The comparison of different contributions to the $\mwb$ distribution
is more clearly shown in  fig.\ref{fig:wtb6}, for $\mt = 150$ GeV.
In fig.\ref{fig:wtb6}a, one can see that below the \tt\ threshold
 the $\mwb$ distribution is dominated by the flatter virtual-$W$
contribution. Note that even above the \tt\ threshold
(\cf\ fig.\ref{fig:wtb6}b, where $\rs = 500$ GeV), when the bulk of cross
section is concentrated at
$\mwb \simeq \mt$, single top production through the virtual-$W$
contribution gives a predominant contribution to the \proc  cross section
 in the large $Wb$ invariant  mass region. Indeed, the virtual-$W$
tail gets about an order of magnitude larger than the $tt^*$ one for
$\mwb \simeq 2\mt$ in fig.\ref{fig:wtb6}b. Anyway, the corresponding
rates will be hardly observable with the foreseen NLC luminosity.

\begin{figure}

\setlength{\unitlength}{1truecm}

\begin{picture}(12.0,12.5)
 \put(-0.5,-3.5){\special{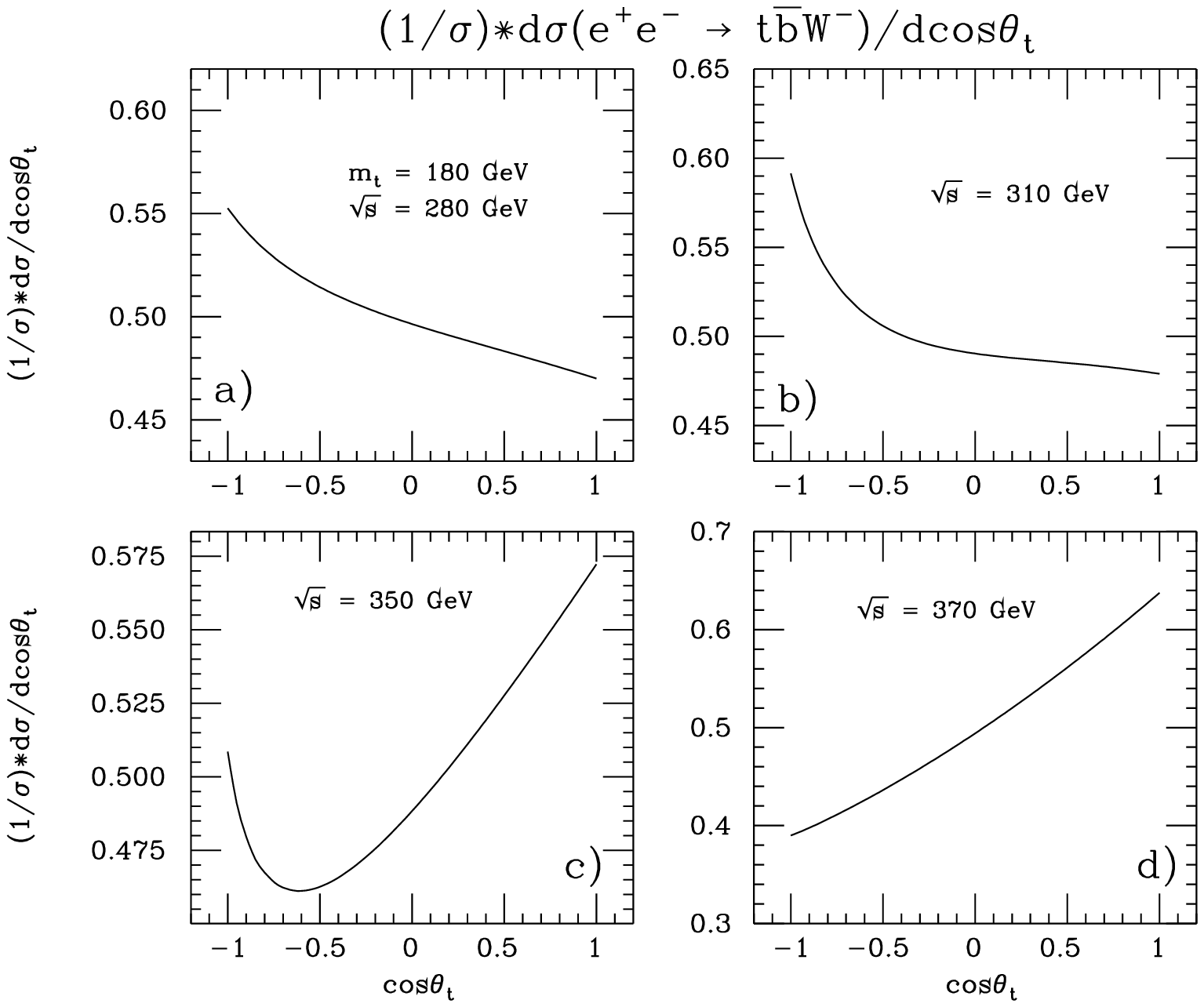}}
\end{picture}

\caption{\rm Angular distribution for the top quark in \protect\proc\
         with respect to the  \protect\elm\ beam direction, in the case
	 $\protect\mt = 180$ GeV and: \protect\\
	 a) $\protect\rs=280$ GeV; b) $\protect\rs=310$ GeV;
	 c) $\protect\rs=350$ GeV; d) $\protect\rs=370$ GeV.}

\protect\label{fig:wtb7}

\end{figure}

An analogous transition from the ``$WW^*$ dominance'' to the
``$tt^*$ dominance'' can be observed in the angular distribution
for the top quark in \protect\proc\  with respect to the \elm\ beam
direction.
 In fig.\ref{fig:wtb7}, we set $\mt = 180$ GeV and
$\rs = 280, \; 310, \; 350, \; 370$ GeV (figs.\ref{fig:wtb7}a,b,c,d
respectively).
By varying $\rs$, one can see that two different behaviours compete.
 At low $\rs$, a collinear $t$ to the \elp\ beam is favoured and
 prevails in the region well below the \tt\ threshold (fig.\ref{fig:wtb7}a,b).
On the other hand, as $\rs$ approaches $2 \mt$, $d\sigma/d\!\costop$
is enhanced in the positive
$\costop$ region, and top quarks that are collinear to the \elm\ are dominant.

\begin{figure}

\setlength{\unitlength}{1truecm}

\begin{picture}(12.0,18.0)
 \put(0.3,4.5){\special{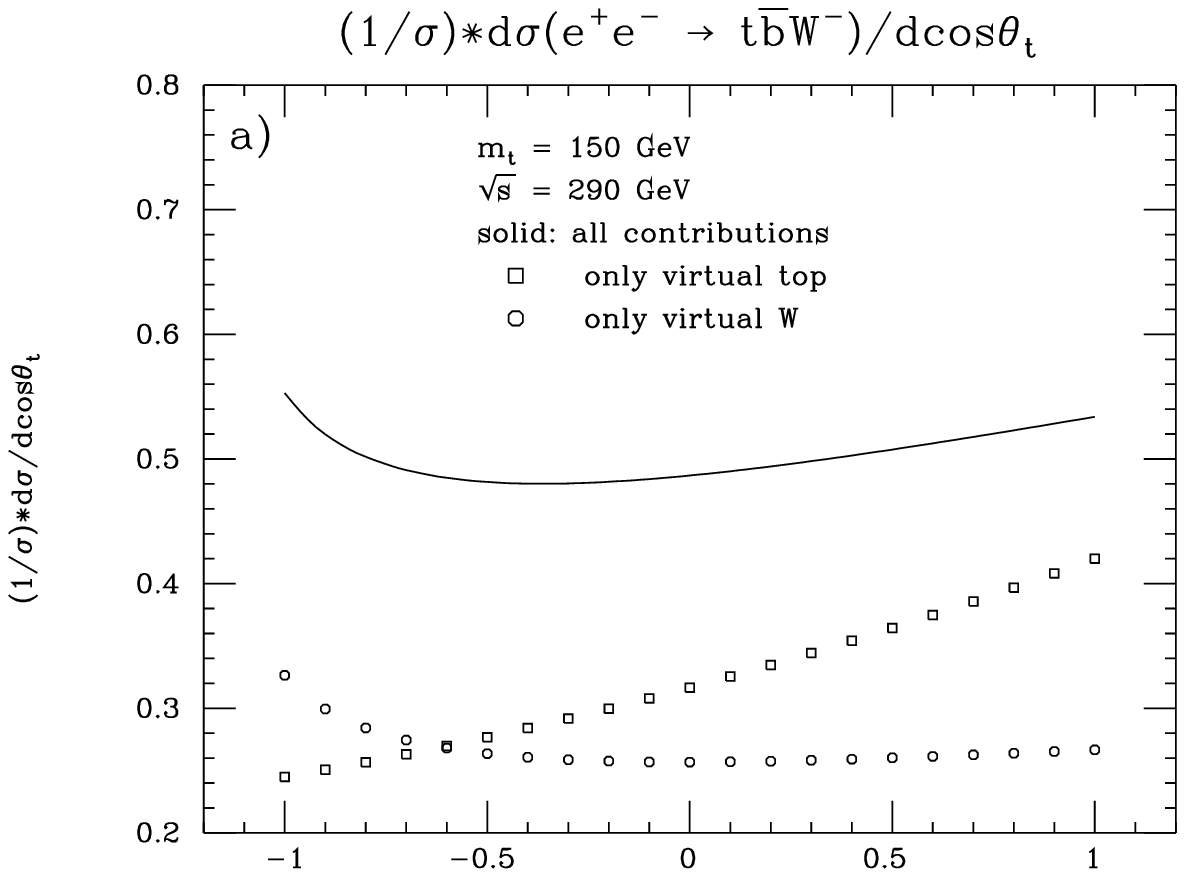}}
 \put(0.3,-4.5){\special{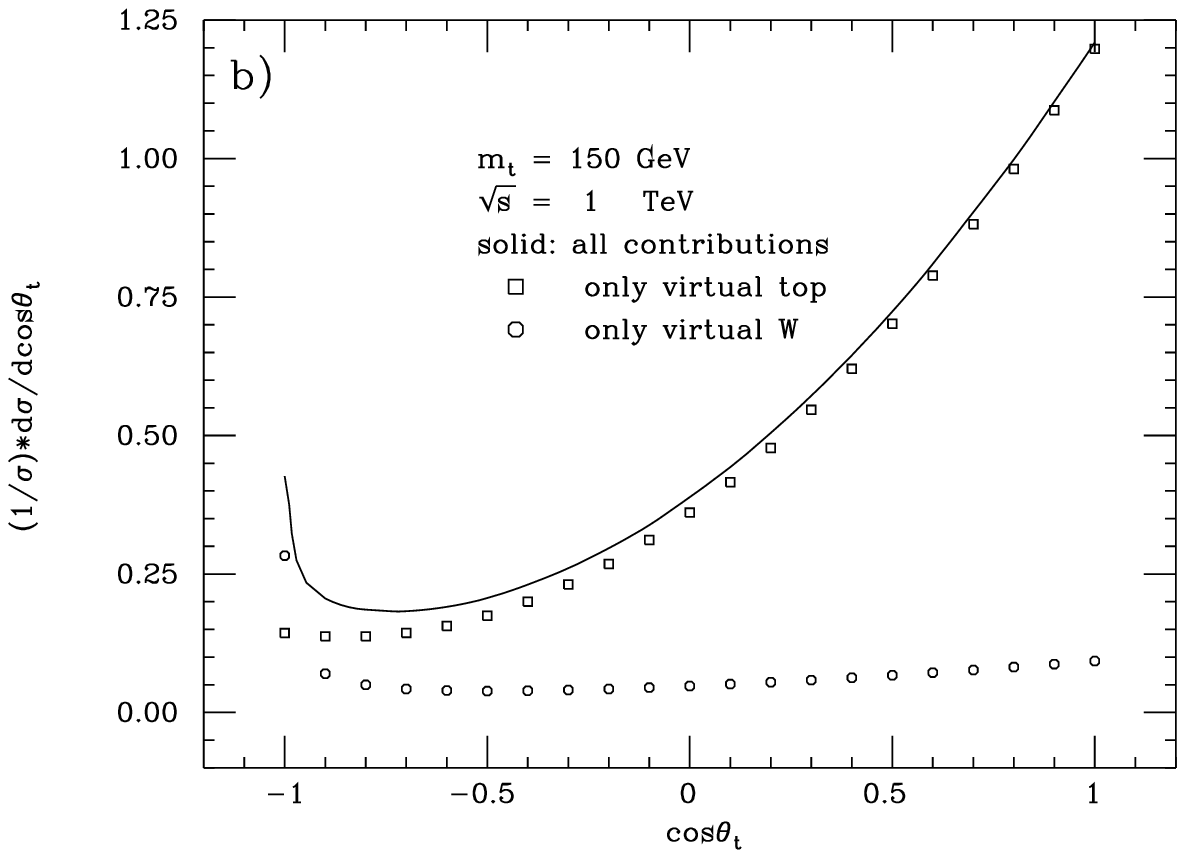}}
\end{picture}

\caption{\rm Various contributions to $\costop$ distribution in the case
	 $\mt = 150$ GeV. \protect\\
	 a) $\protect\rs = 290$ GeV, b) $\protect\rs = 1$ TeV.
	 Notations are  as in fig.\protect\ref{fig:wtb6}.}

\protect\label{fig:wtb8}

\end{figure}

 Fig.\ref{fig:wtb8} shows that
the virtual-$W$ contribution is responsible for the first effect while
virtual-$t$ diagrams produce rather forward-peaked top quarks
 along the \elm\ beam.
In fact, what one observes in  $d\sigma/d\!\costop$ reflects the
differences in the angular distributions for the  ``parent'' processes
\procww\ and \proctt\ .
In \procww, the \Wp\ is produced
preferentially along the \elp\ direction (see, \eg, ref.\cite{barger87}).
The top quark produced through the virtual-$W$ diagrams
in \proc\ and coming from the $W^*$ decay, due to its heaviness, well
reflects the would-be final \Wp\ direction in \procww .
Hence, the top from $WW^*$ will be preferentially
along the \elp\ direction too.
On the contrary, the $t$ quark produced in the $tt^*$ channel
reflects the $t$ angular distribution in \proctt, that is always
peaked toward the \elm\ direction \cite{bern91}.

In fig.\ref{fig:wtb8}, the case $\mt = 150$ GeV is considered in the region
just below the threshold ($\rs = 290$ GeV, fig.\ref{fig:wtb8}a),
where the $WW^*$ and $tt^*$ contributions are comparable,
and at very high \cm\ energy ($\rs = 1$ TeV, fig.\ref{fig:wtb8}b).
In the latter case, it is possible to distinguish a small peak at $\costop
= -1$ due to the relative enhancement at high $\rs$ of the
virtual-$W$ contribution (\cf\ fig.\ref{fig:wtb4}).

\section{Summary and conclusions}
\label{sec:conc}

 In this paper, we have studied  single top production
in \epem\ collisions through the processes \proc, $\bar{t} b W^{\sss +}$,
considering  aside the $tt^*$ channel, that is relevant for top
pair production, $WW^*$ and $bb^*$ channels, with
$W^*\rar tb$ and $b^*\rar tW$ respectively. Since present limits on $\mt$
imply that the top is considerably heavier than the
$W$ vector boson, this process will be of relevance for top production
at Next Linear \epem\ Colliders  operating at a \cm\
energy below the \proctt\ threshold. This, for instance, will be the case
if the top is heavier than 150 GeV for a NLC  with  $\rs \simeq 300$
GeV or less. We found that, assuming an integrated luminosity of
about 10 fb\r{-1}, production rates are sufficient to detect a top signal
below the \tt\ threshold for a large fraction of the kinematically allowed
energy window $\mt + \mw + \mb < \rs < 2 \mt$.
In particular, considering \epem\ collisions at 240 $\ltap \rs \ltap
400$ GeV, by varying $\mt$ from 150 GeV up to 200 GeV, we
get cross sections from 4 to 9 fb for $\rs = 2 \mt - 20$ GeV
and from 0.9 to 4 fb for $\rs = 2 \mt - 40$ GeV.
For 180 $< \mt <$ 200 GeV,  one still obtains cross sections of the
order of 1 fb or more for $\rs = 2 \mt - 60$ GeV.
The above rates include both the single $t$ and $\bar{t}$ production.

Invariant $Wb$ mass distributions and top angular distributions
have been studied and kinematical differences with respect to the \tt\ pair
production above threshold have been stressed.

The relative importance of the  $tt^*$, $WW^*$ and $bb^*$
contributions has been extensively studied both for cross sections
and kinematical distributions . We found that the virtual-$W$ exchange
is predominant in the total cross section for
$\protect\rs \ltap 2 \mt - 20$ GeV.
For instance, if $\mt = 200$ GeV, $WW^*$ contribution is more
than one order of
magnitude larger than the $tt^*$ one for $\rs \ltap 320$ GeV.
Consequently, also the kinematical features of the final states
are governed by the $WW^*$ channel well below the \tt\ threshold.
We have a rather broad $\mwb$ distribution that reflects the non-resonant
structure of the $Wb$ final state, as opposed to the $tt^*$ case.
Furthermore, the top angular distribution recalls the
\Wp\ one  in the process \procww\, that is quite opposite to the $t$
distribution coming from the $tt^*$ contribution.

We also found  that, due basically to the different behaviour
of the \proctt\ and \procww\ cross sections versus $\rs$,
single top production mediated by a virtual $W$ can be non-negligible
far above the \tt\ threshold (\cf\ fig.\ref{fig:wtb4}), although
$tt^*$ and $WW^*$ interference
effects tend to cancel the eventual increase in the total top production rate.
However, for $\rs \simeq 1$-2 TeV and sufficiently large luminosity,
some $WW^*$ effect could be detectable by studying
distributions in particular ranges of kinematical variables
(\cf\ fig.\ref{fig:wtb8}b).

Regarding the $bb^*$ contribution
to the \proc\ cross section, this is found to be
always negligible.

 As far as detection of a \proc\ signal is concerned,
as a result of the fast top decay, one  observes a
$W^{\sss +}W^{\sss -}\bar{b}b$ final state.
Hence, the study of single top production rates below the \tt\ threshold are
interesting not only by itself.
A detailed knowledge of this process also leads to a better determination of
the  possible background for any process of moderate cross section involving
multi-vector boson final states.

\vspace{.5cm}
\noindent
{\large \bf Acknowledgements}

We are indebted to S.~Ilyin and A.~Pukhov for useful
discussions on the \mbox{{\sc CompHEP}} software.



\begin{thebibliography}{999}

\bibitem{hawaii}        Proc. of the 2\r{nd}\ International Workshop on
                        ``Physics and Experiments with Linear \epem\
                        Colliders'', Waikoloa, Hawaii, April 26-30, 1993
                        (in preparation).

\bibitem{hamburg}       Proc. of the 2\r{nd}\ Workshop
                        ``\elp\elm\ Collisions at 500 GeV: The Physics
                        Potential'', Munich, Annecy, Hamburg,
                        Nov 20, 1992 - Apr 3, 1993, ed. P.M.~Zerwas,
                        DESY 92-123C (1994) (in preparation).

\bibitem{zerwas}      P.M.~Zerwas, in the
                      Proc. of the 1\r{st}\ International Workshop on
                      ``Physics and Experiments with Linear \epem\
                      Colliders'', Saariselk\"a, Finland, 9-14 Sep 1991,
                      eds. R.~Orava, P.~Eerola, M.~Nordberg, p.165
                      and references therein.

\bibitem{top}        talks by P.~Tipton (CDF collaboration) and
                     N.~Hadley (D0 collaboration) at the
                     XVI Int. Symp. on Lepton-Photon Interactions,
                     Cornell University, Ithaca, N.Y., U.S.A.,
                     August 10-15, 1993.

\bibitem{lefrancois} plenary talk by J.~Lefran\c{c}ois at the Int. Europhysics
                     Conf. on High Energy Physics, Marseille,
                     July 22-28, 1993.

\bibitem{bern91}        W.~Bernreuther \etal, in the Proc. of the
                        1\r{st}\ Workshop ``\elp\elm\ Collisions at 500 GeV:
                        The Physics Potential'', Munich, Annecy, Hamburg,
                        Feb 4-Sep 3, 1991, ed. P.M.~Zerwas, DESY 92-123A
                        (1992), p.255.

\bibitem{wwz}           V.~Barger, T.~Han and R.J.N.~Phillips,
                        \prev{D39}{89}{146}.

\bibitem{mele93}        M.~Baillargeon, F.~Boudjema, F.~Cuypers,
                        E.~Gabrielli and B.~Mele, preprint
                        CERN-TH-6932-93, ENSLAPP-A-425-93, LMU-10-93,
                        June 1993.

\bibitem{majerotto}     A.~Bartl and W.~Majerotto,
                        in the Proc. of the 2\r{nd}\ Workshop
                        ``\elp\elm\ Collisions at 500 GeV: The Physics
                        Potential'', Munich, Annecy, Hamburg,
                        Nov 20, 1992 - Apr 3, 1993, ed. P.M.~Zerwas,
                        DESY 92-123C (1994) (in preparation).

\bibitem{katuya}        M.~Katuya, J.~Morishita, T.~Munehisa and
                        Y.~Shimizu, \ptp{75}{86}{92}.

\bibitem{panella}       O.~Panella, G.~Pancheri and Y.N.~Srivastava,
                        \pl{B318}{93}{241}.

\bibitem{raidal}        M.~Raidal and R.~Vuopionper\"a, \pl{B318}{93}{237}.

\bibitem{comphep}       {\sc CompHEP}, {\it a System for Computations in
                        High Energy Physics}, (1990-93), V.~Savrin and
                        S.~Ilyin (Sc. Coord.), A.~Pukhov (Comp.), E.~Boos
                        and M.~Dubinin (Phys.), Moscow State Univ., Inst.
                        for Nucl. Phys., Symb. Comp. Lab.; \\
                        E.~Boos \etal, ``New Computing Techniques in
                        Phys. Research II'', Proc. of the 2\r{nd}\ Int.
Workshop
                        on Software Engineering, Artificial Intelligence
                        and Expert Systems in High Energy and Nuclear Physics,
                        ed. D.~\mbox{Perret-Gallix} World Scientific, 1992,
                        p.665.

\bibitem{vegas}         G.P.~Lepage, \jcp{27}{78}{192}.

\bibitem{kuhn}          M.~Je\.{z}abek and J.H.~K\"uhn, \np{B314}{89}{1};
			M.~Je\.{z}abek and J.H.~K\"uhn, preprint TTP-93-4;
                        A.~Denner and T.~Sack, \np{B358}{91}{46};
			G.~Eilam, R.R.~Mendel, R.~Migneron and A.~Soni,
                        \prl{66}{91}{3105}.

\bibitem{kuhn81}        J.H.~K\"uhn, \app{B12}{81}{347};
                        J.H.~K\"uhn, \apa{Suppl.XXIV}{82}{203};
                        I.~Bigi, Y.L.~Dokshitzer, V.A.~Khoze, J.H.~K\"uhn
                        and P.M.~Zerwas, \pl{B181}{86}{157}.

\bibitem{barger87}      V.~Barger and R.J.N.~Phillips, {\it ``Collider
                        Physics''}, ed. David Pines, 1987, p.268.

\end{thebibliography}
\end{document}